\documentclass[prx,twocolumn, amsmath,amssymb,
reprint,%
author-year,%
author-numerical,%
dvipdfmx
]{revtex4}

\usepackage{graphicx}
\usepackage{bm}
\usepackage{color}

\begin{document}


\title{Theory of dynamical stability for two- and three-dimensional Lennard-Jones crystals}

\author{Shota Ono}
\email{shota\_o@gifu-u.ac.jp}
\author{Tasuku Ito}
\affiliation{Department of Electrical, Electronic and Computer Engineering, Gifu University, Gifu 501-1193, Japan}

\begin{abstract}
The dynamical stability of three-dimensional (3D) Lennard-Jones (LJ) crystals has been studied for many years. The face-centered-cubic and hexagonal close packed structures are dynamically stable, while the body-centered cubic structure is stable only for long range LJ potentials that are characterized by relatively small integer pairs $(m,n)$. Here, we study the dynamical stability of two-dimensional (2D) LJ crystals, where the planar hexagonal, the buckled honeycomb, and the buckled square structures are assumed. We demonstrate that the stability property of 2D and 3D LJ crystals can be classified into four groups depending on $(m,n)$. The instabilities of the planar hexagonal, the buckled square, and the body-centered cubic structures are investigated within analytical expressions. The structure-stability relationship between the LJ crystals and the elemental metals in the periodic table is also discussed.  
\end{abstract}

\maketitle

\section{Introduction}
Lattice vibration is of importance to understand the dynamical properties of solids. The crystal lattice is dynamically stable if the vibrational frequency is real over the entire Brillouin zone, while if a specific vibrational mode has an imaginary frequency, the crystal lattice is unstable against such a mode \cite{grimvall}. Due to the development of first-principles approach such as density-functional theory (DFT) \cite{KS}, the dynamical stability has been studied in a variety of materials \cite{kawazoe,dfpt,togo}. For simple metals, the face-centered cubic (FCC) metals in the hexagonal close packed (HCP) structure and the HCP metals in the FCC structure are dynamically stable, i.e., metastable state. However, the FCC metals in the body-centered cubic (BCC) structure and vice versa are, in general, dynamically unstable \cite{grimvall}. Also metastable phases have attracted significant interest in recent years due to their unconventional properties \cite{huang,schonecker}. 

\begin{figure}[bb]
\center
\includegraphics[scale=0.43]{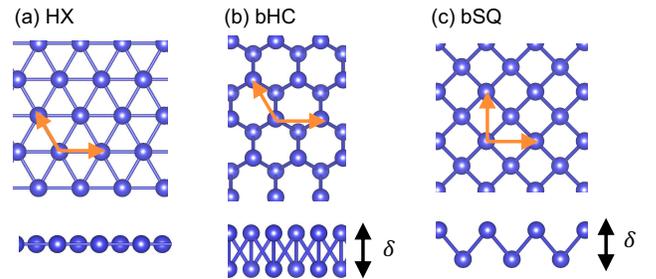}
\caption{The top and side views of the 2D lattice structure: (a) HX, (b) bHC, and (c) bSQ structures. The primitive lattice vectors are indicated by arrows in the top view. $\delta$ is the buckling height.  } \label{fig1} 
\end{figure}

Recently, one of the authors has studied the dynamical stability of two-dimensional (2D) layers of simple metals using the first-principles approach \cite{ono2020_1,ono2020_2}. We have considered three-types of 2D lattice structure: The hexagonal (HX), the buckled honeycomb (bHC), and the buckled square (bSQ) structures, as shown in Fig.~\ref{fig1} \cite{ono2020_2}. It has been proposed that (i) If the HX structure is dynamically stable, the FCC and/or HCP structures are also dynamically stable; (ii) If the bHC structure is dynamically stable only, the BCC structure is dynamically stable; (iii) If the bSQ structure is dynamically stable only, the HCP structure is dynamically stable; and (iv) If the bHC and bSQ structures are dynamically stable, the HCP or FCC structures are stable depending on the group in the periodic table. It is interesting to study the structure stability relationship between 2D and 3D systems by using more simple models that enable us to derive the vibrational frequencies at some symmetry points analytically. 

In this paper, we focus on the Lennard-Jones (LJ) crystals and investigate how the dynamical stabilities in the 2D structures (HX, bHC, and bSQ) and 3D structures (FCC, BCC, and HCP) are correlated with each other. We categorize the stability property into four groups (Fig.~\ref{fig2} below) and discuss the stability relationship between the LJ crystals and the elemental metals in the periodic table. We also provide analytical expressions for studying the dynamical stability of the HX, bSQ, and BCC LJ crystals. 

Among many models, the LJ potential is one of the simplest model for describing the interatomic forces between atoms in materials. The LJ potential is a central potential that depends on the interatomic distance $R$ only and can describe the dynamics of rare gas solids \cite{mermin}. For positive integer pairs $(m,n)$, the generalized LJ potential is given by 
\begin{eqnarray}
 \phi (R) = \frac{C_m}{R^{m}} - \frac{C_n}{R^{n}}
 \label{eq:LJ}
\end{eqnarray}
with $C_m$ and $C_n$ being the positive values. The first and second terms in Eq.~(\ref{eq:LJ}) describe the repulsive and attractive potential between atoms. The inequality $m>n$ must hold since an atom cannot penetrate into the volume of other atoms. It has been known that the FCC lattice is dynamically stable for all $m>n$, while the BCC lattice is unstable except for small $m$ and $n$ \cite{born,wallace}. Modified LJ potentials taking the Friedel oscillations into account have been used to study the stability of alloys \cite{mihalkovic}. The Morse potential that is also a central potential has been used to model the chemical bonds in alloys \cite{alsalmi}. For an accurate description beyond the central potential approximation, many-body corrections are required, as implemented in such as the embedded atom method \cite{finnis,foiles} and DFT \cite{KS}. Therefore, we expect that the investigation regarding the LJ potentials would provide a simple understanding for the dynamical stability of solids. 

\section{Theory}
\subsection{Basic concepts}
Throughout this paper, we use the term of ``atom'' to indicate the LJ particle. The total potential energy for $N$-atom systems is defined by
\begin{eqnarray}
 V = \frac{1}{2}\sum_{i=1}^{N}\sum_{j(\ne i)}^{N} \phi(R_{ij}),
 \label{eq:vtot}
\end{eqnarray}
where $R_{ij}$ is the interatomic distance between atoms $i$ and $j$. The factor of $1/2$ accounts for the double counting of the summation. 

Within the harmonic approximation \cite{mermin}, the lattice dynamics of solids is determined by 
\begin{eqnarray}
 M \frac{d^2 u_{\alpha}(s,\bm{l})}{dt^2} 
 = - \sum_{\alpha's'\bm{l}'} D_{ss'}^{\alpha\alpha'}(\bm{l},\bm{l}')
 u_{\alpha'}(s',\bm{l}'),
 \label{eq:newton}
\end{eqnarray}
with the particle mass $M$ and the $\alpha(=x,y,z)$ component of the displacement $u_{\alpha}(s,\bm{l})$ for $s$th atom in a unit cell characterized by the lattice vector $\bm{l}$. The force constant matrix is defined by
\begin{eqnarray}
D_{ss'}^{\alpha\alpha'}(\bm{l},\bm{l}') 
= \frac{\partial^2 V}{\partial R_{\alpha}(s,\bm{l}) \partial R_{\alpha'}(s',\bm{l}') }\Big\vert_0,
\end{eqnarray}
where the derivative is evaluated at the equilibrium atom positions $R_{\alpha}(s,\bm{l})$. Assuming the plane wave solution
\begin{eqnarray}
u_{\alpha}(s,\bm{l}) = \epsilon_{\alpha, s}(\bm{q}) e^{i\bm{q}\cdot \bm{R}(s,\bm{l})} e^{- i\omega t} 
\end{eqnarray}
with the frequency $\omega$ and the $\alpha$-component of the polarization vector $\epsilon_{\alpha, s}(\bm{q})$ for the wavevector $\bm{q}$, one obtains
\begin{eqnarray}
\omega^2 \epsilon_{\alpha, s}(\bm{q}) = 
\sum_{\alpha' s'} \tilde{D}_{ss'}^{\alpha\alpha'}(\bm{q}) \epsilon_{\alpha', s'}(\bm{q}),
\end{eqnarray}
where $\tilde{D}$ is the dynamical matrix given by
\begin{eqnarray}
 \tilde{D}_{ss'}^{\alpha\alpha'}(\bm{q}) 
 = \frac{1}{M}\sum_{\bm{l}} D_{ss'}^{\alpha\alpha'}(\bm{0},\bm{l}) 
 e^{i\bm{q}\cdot \left[ \bm{R}(s',\bm{l}) - \bm{R}(s,\bm{0})\right]}. 
 \label{eq:dyn}
\end{eqnarray}
The acoustic sum rule, that is, $\sum_{s'\bm{l}} D_{ss'}^{\alpha\alpha'}(\bm{0},\bm{l}) =0$, is imposed by considering the translational symmetry of the crystal. When there is only one atom in the unit cell, Eq.~(\ref{eq:dyn}) is written as \cite{mermin}
\begin{eqnarray}
 \tilde{D}^{\alpha\alpha'}(\bm{q}) 
 &=& \frac{2}{M}\sum_{\bm{l}}{}^{'} \left[ A(L)\delta_{\alpha\alpha'} 
+ B(L) (\hat{l}\hat{l})_{\alpha\alpha'}
\right]
\nonumber\\
&\times&
\sin^2\left( \frac{\bm{q}\cdot \bm{l}}{2} \right)
 \label{eq:dyn2}
\end{eqnarray}
with $L=\vert\bm{l}\vert$, $A(L)=\phi'(L)/L$, and $B(L)=\phi''(L) - \phi'(L)/L$, where the prime denotes the derivative with respect to $R$. $(\hat{l}\hat{l})_{\alpha\alpha'}$ is the dyadic defined as $l_\alpha l_{\alpha'}/L^2$. The summation in Eq.~(\ref{eq:dyn2}) is taken over the lattice vectors except $\bm{l}=0$. If the eigenvalue $\lambda_j (\bm{q})$ of $\tilde{D}$ for $\bm{q}$ and $j=1,\cdots,3n_a$ ($n_a$ being the number of atoms in a unit cell) is positive (negative), the $N$-atom system is dynamically stable (unstable) against the vibrational mode ($\bm{q},j$). 

\subsection{LJ potential optimization}
In order to determine the parameters of the LJ potential given by Eq.~(\ref{eq:LJ}), we first consider the HX structure because the HX structure can be building blocks for constructing other structures such as FCC, HCP, and bHC structures. The primitive lattice vectors of the HX structure are given by 
\begin{eqnarray}
 \bm{a}_1 = (a,0,0), \ \ 
 \bm{a}_2 = \left(-\frac{a}{2},\frac{\sqrt{3}a}{2},0\right), \ \
 \bm{a}_3 = (0,0,c) 
 \label{eq:primitive_HX}
\end{eqnarray}
with the lattice constant $a$. The interlayer distance $c$ is taken to be much larger than $a$ in order to study the lattice dynamics of the isolated thin films. We set the total energy per atom $V/N$ and the lattice parameter $a$ to be $-E_0$ and $a_0$, respectively, providing the following simultaneous equations
\begin{eqnarray}
 \left(
 \begin{array}{c}
 -E_0 \\
 0
 \end{array}
 \right)
 = 
 \left(
 \begin{array}{cc}
 J_m  & - J_n \\
 m J_m & - n J_n \\
\end{array}
\right)
\left(
 \begin{array}{c}
 C_m \\
 C_n
 \end{array}
 \right),
\end{eqnarray}
where $J_m$ for an integer $m$ is defined as 
\begin{eqnarray}
 J_m = \frac{1}{2} \sum_{j\ne i} \frac{1}{R_{ij}^{m}}
\end{eqnarray}
with the minimum of $R_{ij}$ being $a_0$. The LJ potential parameters $C_m$ and $C_n$ are then determined by
\begin{eqnarray}
 C_m = \frac{nE_0}{(m-n)J_m}, \ \ \ C_n = \frac{mE_0}{(m-n)J_n}. 
 \label{eq:LJcoef}
\end{eqnarray}

By using Eq.~(\ref{eq:LJcoef}), we perform the structure optimization for the bHC and bSQ structures. The bHC structure is formed from the primitive lattice vectors given by Eq.~(\ref{eq:primitive_HX}), with a two-point basis: the positions of the particle 1 and 2 are given by $(0, 0, 0)$ and $(0, a/\sqrt{3},\delta)$, respectively, where $\delta$ is the buckling height. The bSQ structure is created from the primitive lattice vectors 
\begin{eqnarray}
 \bm{a}_1 = (a,0,0), \ \ \bm{a}_2 = (0,a,0), \ \ \bm{a}_3 = (0,0,c)
\end{eqnarray}
and from a two-point basis given by $(0, 0, 0)$ and $(a/2, a/2, \delta)$. The optimization of $a$ and $\delta$ for bHC and bSQ structures is performed by using the Newton's method. For the cases of FCC, BCC, and HCP, the size of $a$ is optimized only. The ratio of $c/a$ in the HCP structure is fixed to $\sqrt{8/3}$. 

The LJ potential is studied for the set of integers, $(m,n)$, satisfying $3\le n \le 8$ and $n < m \le 12$. Since $\phi (R)$ in Eq.~(\ref{eq:LJ}) goes to zero monotonically when $R\rightarrow \infty$, we set the cutoff radius $L_{\rm c}$ for $\phi (R)$ to be 20$a_0$. This value is large enough to study the stability property of LJ crystals. In fact, although an increase in $L_{\rm c}$ up to 30$a_0$ can produce lower total energy per atom, the stability property (Fig.~\ref{fig2} below) is the same as that in the case of $L_{\rm c}=20a_0$.

In the present study, the parameters $C_m$ and $C_n$ in Eq.~(\ref{eq:LJ}) are optimized for the HX structure. It is noteworthy that even when these are optimized for the FCC structure, the main result in the present study, Fig.~\ref{fig2} below, does not change. This indicates that the structure-stability property in the LJ crystals is determined by $(m,n)$ only. 

\begin{figure}[tt]
\center
\includegraphics[scale=0.48]{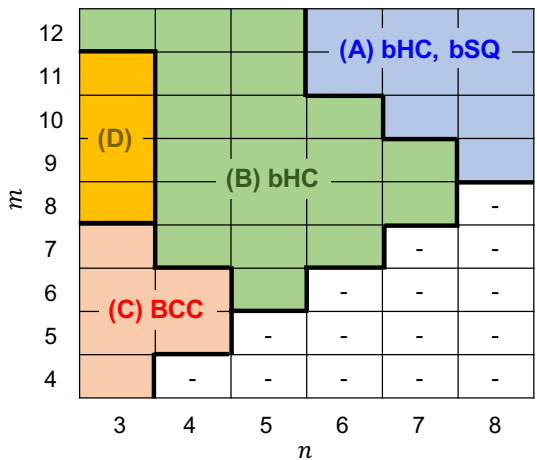}
\caption{The stability property for the $(m,n)$-LJ crystals in the BCC, bHC, and bSQ structures. For the group D, i.e., $(m,n)=(8,3), (9,3), (10,3), (11,3)$, the LJ crystals are unstable. } \label{fig2} 
\end{figure}

\begin{figure}[tt]
\center
\includegraphics[scale=0.4]{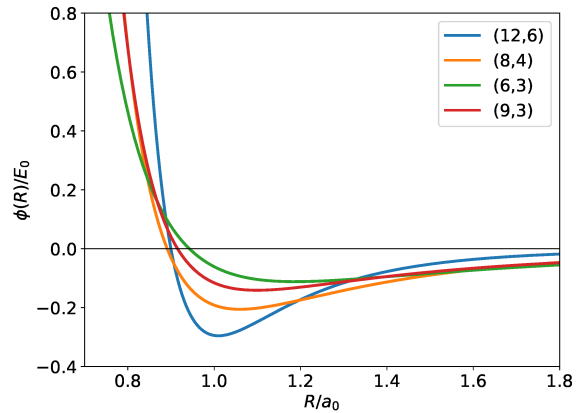}
\caption{The $R$-dependence of the (12,6), (8,4), (6,3), and (9,3)-LJ potentials.  } \label{fig3} 
\end{figure}

\begin{table*}
\begin{center}
\caption{$R_0$ in the LJ potential and the optimized values ($E$, $a$, and $\delta$) for the bHC and bSQ structures. $R_0$, $a$, and $\delta$ are in units of $a_0$, whereas $E$ is in units of $E_0$. }
{
\begin{tabular}{lccccccc}\hline
   & $R_0$ \hspace{5mm}  & $E_{\rm bHC}$ \hspace{5mm}  & $a_{\rm bHC}$ \hspace{5mm}   &   $\delta_{\rm bHC}$ \hspace{5mm}  &  $E_{\rm bSQ}$ \hspace{5mm}  &   $a_{\rm bSQ}$ \hspace{5mm}  &  $\delta_{\rm bSQ}$  \\ \hline
(12,6) in group A \hspace{5mm} & 1.010 \hspace{5mm} & -1.656 \hspace{5mm} & 0.989 \hspace{5mm} & 0.818  \hspace{5mm} & -1.503 \hspace{5mm} & 0.979 \hspace{5mm} & 0.735   \\
(8,4) in group B \hspace{5mm}  & 1.060 \hspace{5mm} & -1.869 	\hspace{5mm} & 0.963 \hspace{5mm} & 0.870 \hspace{5mm} & -1.716 \hspace{5mm} & 0.946 \hspace{5mm} & 0.831   \\
(6,3) in group C \hspace{5mm}  & 1.187 \hspace{5mm} & -2.129 \hspace{5mm} & 0.928 \hspace{5mm} & 0.967 \hspace{5mm} & -1.988 \hspace{5mm} & 0.885 \hspace{5mm} & 0.990   \\
(9,3) in group D \hspace{5mm}  & 1.099 \hspace{5mm} & -1.985 \hspace{5mm} & 0.948 \hspace{5mm} & 0.899 \hspace{5mm} & -1.820 \hspace{5mm} & 0.923 \hspace{5mm} & 0.866   \\
\hline
\end{tabular}
}
\label{table1}
\end{center}
\end{table*}

\begin{table*}
\begin{center}
\caption{The optimized values ($E$ and $a$) for the FCC, BCC, and HCP structures. $E$ and $a$ are in units of $E_0$ and $a_0$, respectively. }
{
\begin{tabular}{lcccccc}\hline
   & $E_{\rm FCC}$ \hspace{5mm}  & $a_{\rm FCC}$ \hspace{5mm}   &   $E_{\rm BCC}$ \hspace{5mm}  &  $a_{\rm BCC}$ \hspace{5mm}  &   $E_{\rm HCP}$ \hspace{5mm}  &  $a_{\rm HCP}$  \\ \hline
(12,6) in group A \hspace{5mm} & -2.546 \hspace{5mm} &1.387 \hspace{5mm} & -2.435 \hspace{5mm} & 1.110 \hspace{5mm} & -2.546 \hspace{5mm} & 0.981  \\
(8,4) in group B \hspace{5mm} & -4.840 \hspace{5mm} & 1.274 \hspace{5mm} & -4.775 \hspace{5mm} & 1.015 \hspace{5mm} & -4.839 \hspace{5mm} & 0.900   \\
(6,3) in group C \hspace{5mm} & -16.961 \hspace{5mm} &1.011 \hspace{5mm} & -16.870 \hspace{5mm} & 0.804 \hspace{5mm} & -16.963 \hspace{5mm} & 0.715 \\
(9,3) in group D \hspace{5mm} & -10.070 	\hspace{5mm} &1.186 \hspace{5mm} & -9.943 \hspace{5mm} & 0.945 \hspace{5mm} & -10.069 \hspace{5mm} & 0.838 \\
\hline
\end{tabular}
}
\label{table2}
\end{center}
\end{table*}

\section{Results and discussion}
The main results are summarized as follows: For any sets of $(m,n)$, the FCC and HCP structures are dynamically stable, which is consistent with the previous studies \cite{born,wallace}. However, the HX structure is unstable, which is due to the zero thickness of this system, as discussed below. The stability properties for the BCC, bHC, and bSQ structures are presented in Fig.~\ref{fig2}. It is categorized into four groups: (A) The bHC and bSQ structures are dynamically stable (blue); (B) The bHC is dynamically stable only (green); (C) The BCC is dynamically stable only (red); and (D) No stable structures are found (orange). 

As an example for each group, we study the $(12,6)$, $(8,4)$, $(6,3)$, and $(9,3)$-LJ crystals in the HX, bHC, bSQ, FCC, BCC, and HCP structures. The optimized parameters for these structures are listed in Table \ref{table1} and \ref{table2}. The $R_0$s satisfying $\phi'(R_0)=0$ are also listed in Table \ref{table1}. Figure \ref{fig3} shows the $R$-dependence of the (12,6), (8,4), (6,3), and (9,3)-LJ potentials. For the case $(m,n)=(12,6)$, the LJ potential takes the minimum at $R=R_0\simeq a_0$ and decays to zero monotonically as $R$ increases, which will create the short range interaction forces between atoms. As the values of $(m,n)$ decrease, the value of $R_0$ shifts to large $R$, as listed in Table \ref{table1}, and the minimum value of the LJ potential becomes shallow. The curvature around $R$ being the lattice parameter is important to understand the instability of the BCC and 2D structures, which will be discussed below. 

\begin{figure}[tt]
\center
\includegraphics[scale=0.35]{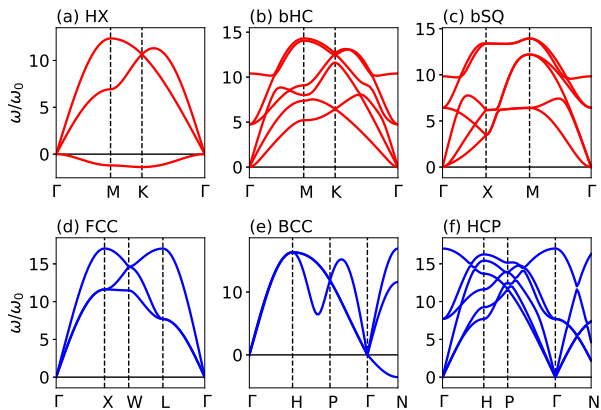}
\caption{The dispersion curves of the (12,6)-LJ crystals for (a) HX, (b) bHC, (c) bSQ, (d) FCC, (e) BCC, and (f) HCP structures.  } \label{fig4} 
\end{figure}

\begin{figure}[tt]
\center
\includegraphics[scale=0.35]{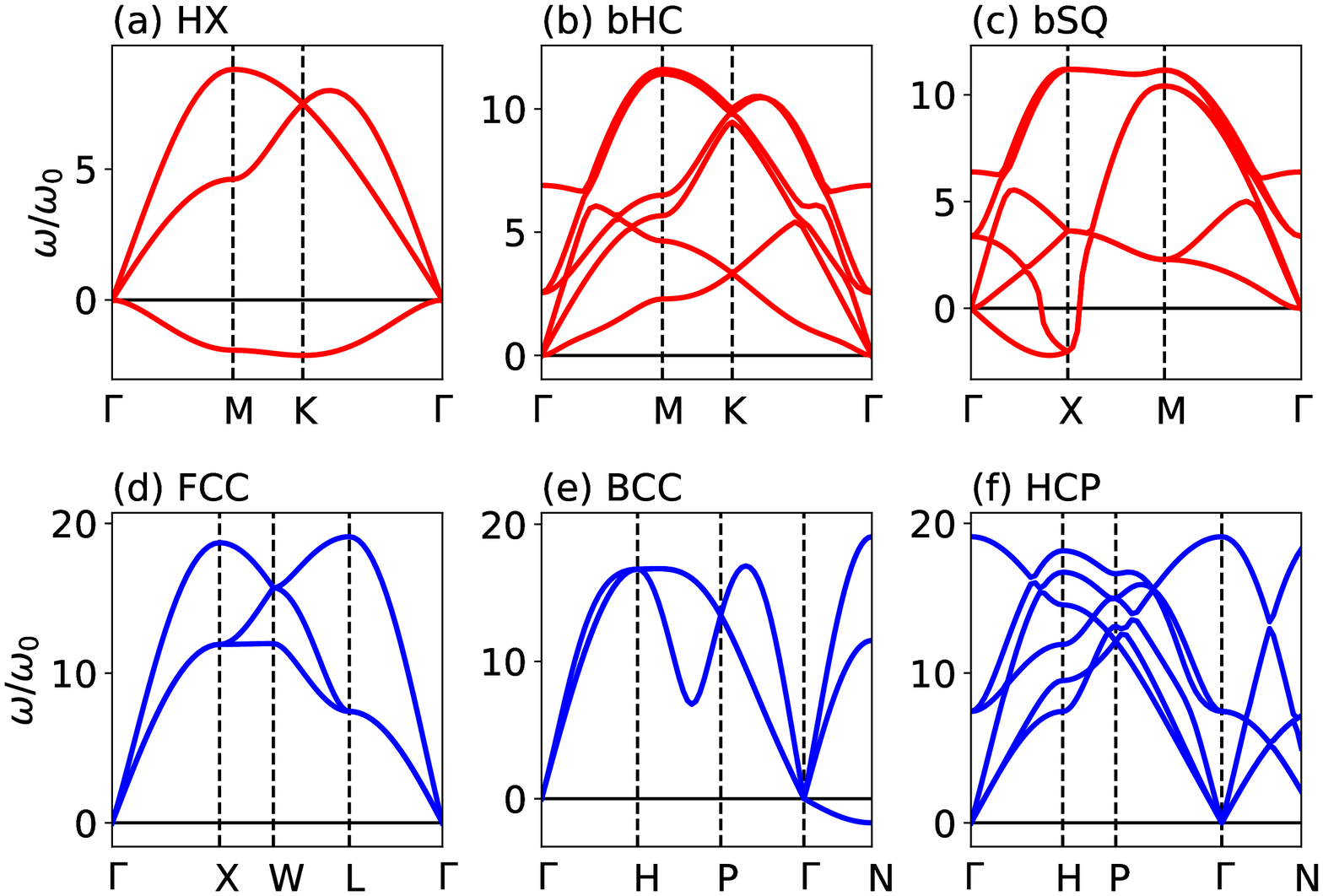}
\caption{Same as Fig.~\ref{fig4} but for the (8,4)-LJ crystals.  } \label{fig5} 
\end{figure}

\begin{figure}[tt]
\center
\includegraphics[scale=0.35]{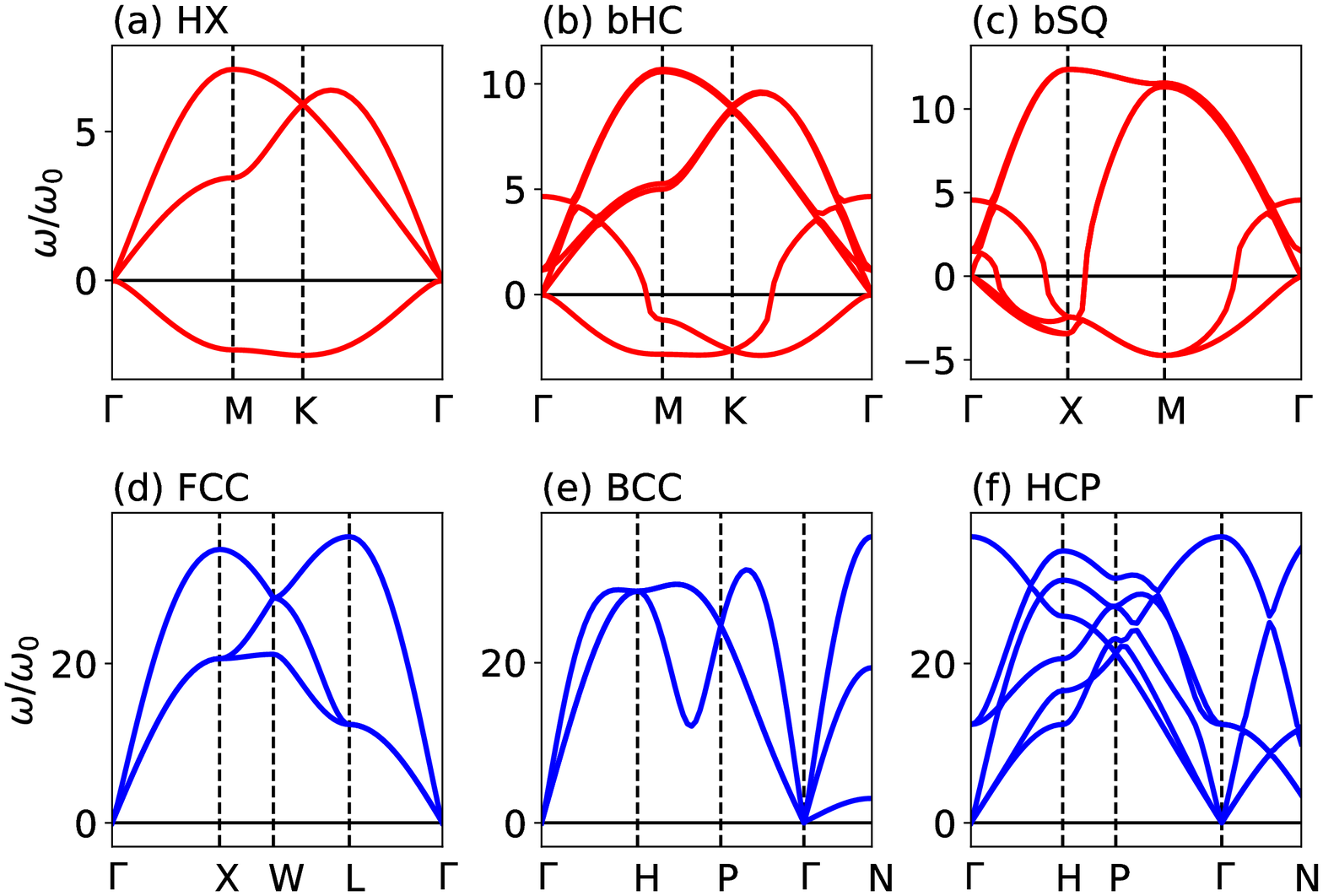}
\caption{Same as Fig.~\ref{fig4} but for the (6,3)-LJ crystals.  } \label{fig6} 
\end{figure}

\begin{figure}[tt]
\center
\includegraphics[scale=0.35]{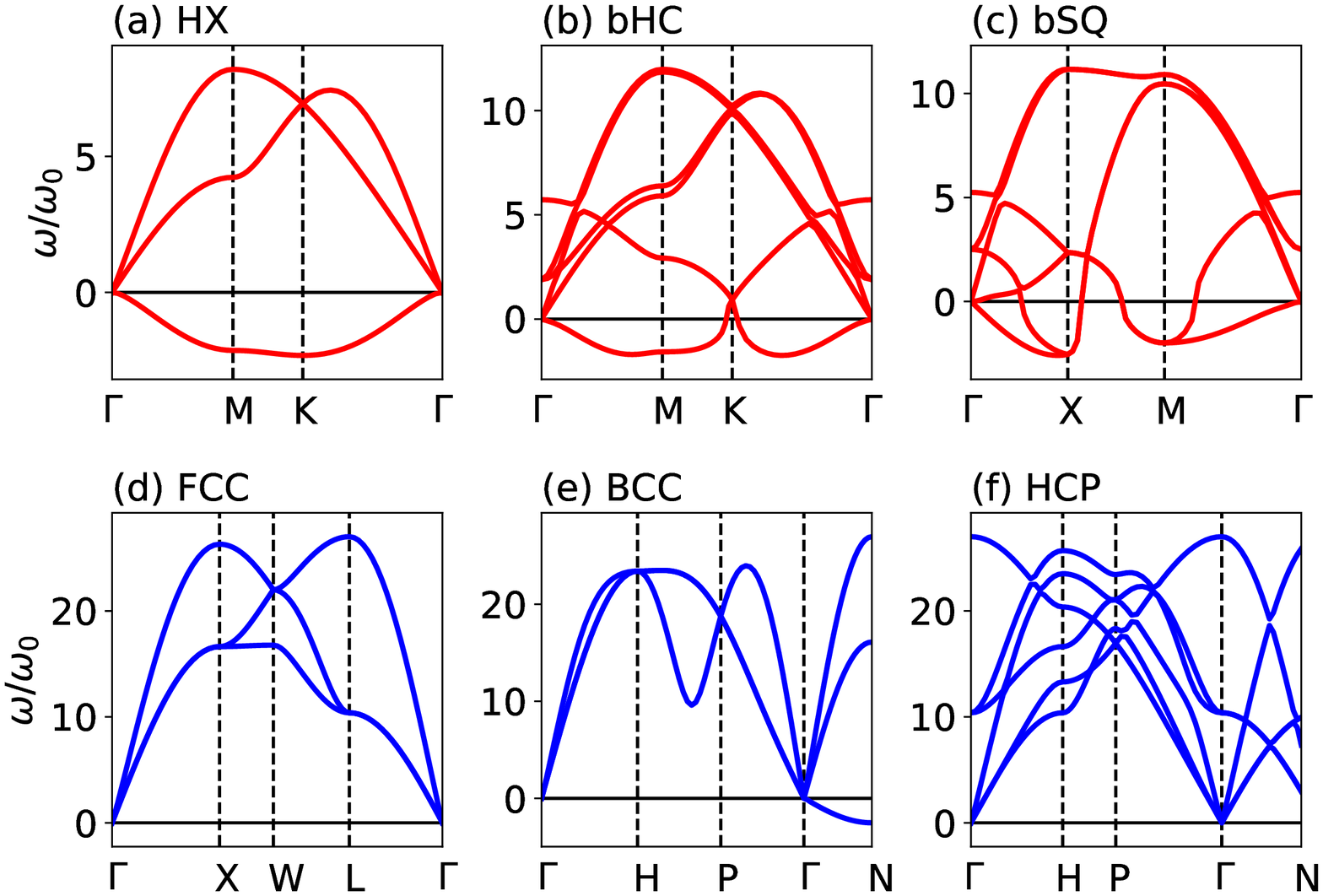}
\caption{Same as Fig.~\ref{fig4} but for the (9,3)-LJ crystals.  } \label{fig7} 
\end{figure}

The dispersion curves for the $(12,6)$, $(8,4)$, $(6,3)$, and $(9,3)$-LJ crystals are shown in Figs.~\ref{fig4}, \ref{fig5}, \ref{fig6} and \ref{fig7}, respectively. The vibrational frequency is expressed in units of $\omega_0 = \sqrt{E_0/(Ma_0^2)}$. The imaginary frequency is represented as negative value of $\omega$. As shown in Figs.~\ref{fig4}(a)-\ref{fig7}(a), the instability of the HX structure is attributed to the appearance of the imaginary frequencies of the out-of-plane (ZA) mode over the Brillouin zone. With Eq.~(\ref{eq:dyn2}), the frequency of the ZA mode for the HX structure is expressed as
\begin{eqnarray}
 \omega(\bm{q})
 = \sqrt{\frac{2}{M}\sum_{\bm{l}}{}^{'} \frac{\phi'(L)}{L}
\sin^2\left( \frac{\bm{q}\cdot \bm{l}}{2} \right)},
\end{eqnarray}
where the summation is taken over the lattice sites having the magnitude of $L = a_0, \sqrt{3}a_0, 2a_0, \cdots$. The dominant contribution is from the sites with $L=a_0$, but $\phi'(a_0)$ is always negative because the minimum of $\phi(R)$ is located at $R=R_0>a_0$, as listed in Table \ref{table1}. The second largest contribution is from the sites with $L = \sqrt{3}a_0$, but the size of $\phi'(\sqrt{3}a_0)(>0)$ is not large enough to stabilize the ZA mode. 

When the buckling is assumed as in bHC and bSQ, the lattice parameter is shortened compared to the HX structure, resulting in lower total energy, as listed in Table \ref{table1}. No imaginary frequencies are observed within the Brillouin zone for the (12,6)-LJ crystals in the bHC and bSQ structure (see Fig.~\ref{fig4}(b) and (c)) and the (8,4)-LJ crystal in the bHC structure (see Fig.~\ref{fig5}(b)). For the (8,4)-LJ crystals, the bSQ structure is unstable against the ZA modes along the $\Gamma$-X direction. The (6,3) and (9,3)-LJ crystals in both the bHC and bSQ are unstable. 

To understand this, we derive analytical expressions for the vibrational frequencies at the point X for the bSQ structure. By considering up to the second nearest neighbor (NN) sites, one obtains an expression for the lowest frequency (doubly degenerate). The derivation is provided in Appendix \ref{app}. The stability condition is given by 
\begin{eqnarray}
 \left[\tilde{A} + B(a_{\rm bSQ})\right]\left[\tilde{A} + \beta B(d_{12})\right] + \alpha \tilde{A} B(d_{12}) > 0
 \label{eq:XbSQ}
\end{eqnarray}
with $\tilde{A}=A(a_{\rm bSQ})+A(d_{12})$, $\alpha = a^2/(4d_{12}^2)$, $\beta = \delta^2/d_{12}^2$, and $d_{12}^{2} = a^2/2 + \delta^2$. For the case of $(m,n)=(12,6)$, the optimized lattice parameter ($a_{\rm bSQ}$) and the interatomic distance between the particle 1 and 2 ($d_{12}$) are almost equal to the location of the potential minimum, that is, $a_{\rm bSQ}\simeq R_0$ and $d_{12}= 1.009a_0 \simeq R_0$, as listed in Table \ref{table1}. This leads to $\tilde{A}\simeq 0$ and $0<B(a_{\rm bSQ}) \simeq B(d_{12})$, so that the inequality in Eq.~(\ref{eq:XbSQ}) is easily satisfied. For the case of $(m,n)=(6,3)$, the size of $a_{\rm bSQ}=0.885a_0$ is smaller than $d_{12} = 1.171 a_0 \simeq R_0$, leading to $A(a_{\rm bSQ}) <0$ (see Fig.~\ref{fig3}). In addition, since the LJ potential is shallow, the magnitude of $B(d_{12})(>0)$ is not so large enough to cancel negative $\tilde{A}$ in the same brackets in Eq.~(\ref{eq:XbSQ}). This gives rise to the instability of the bSQ structure. This is due to the long-range nature of the (6,3)-LJ potential: The size of $a_{\rm bSQ}$ becomes small in order to increase the energy gain, whereas $a_{\rm bSQ}$ is shifted from $R_0$. Unfortunately, it was difficult to derive the stability condition analytically for the bHC structure because the off-diagonal elements in Eq.~(\ref{eq:dyn}) are not zero.

The dynamical stability of the BCC structure shows opposite tendency: When the long and short range LJ potentials are used, the BCC structure is dynamically stable and unstable, respectively. As shown in Figs.~\ref{fig4}(e)-\ref{fig7}(e), the BCC is dynamically stable for the (6,3)-LJ crystal only, otherwise it is unstable against the vibrational modes along $\Gamma$-N direction. The lowest frequency at the point N is expressed by \cite{ono2019_jap,ono2020_jap}
\begin{eqnarray}
 \omega_{\rm N} = \sqrt{\frac{4\left[A(a_1) + A(a_2) \right] + 2B(a_2)}{M}},
\end{eqnarray}
where in Eq.~(\ref{eq:dyn2}) the force constants up to the second NN sites, $a_1 = \sqrt{3}a_{\rm BCC}/2$ and $a_2 = a_{\rm BCC}$, are considered. Since $A(a_1)<0$, the positive value of $B(a_2)$ must be large enough to overcome the negative contribution from $A(a_1)$. When the set $(m,n)=(6,3)$ is used, the size of $a_{\rm BCC}$ is about $0.8a_0$, yielding $a_1 \simeq 0.7a_0$ and $a_2 = 0.8a_0$, respectively. Since these are less than $R_0=1.187a_0$, a relation $B (a_2)\gg \vert A (a_1)\vert$ holds, so that the BCC becomes dynamically stable. 

Finally, we discuss the stability relationship between the LJ crystals and the elemental metals in the periodic table. In the previous study \cite{ono2020_2}, it has been shown that most transition metals in the bHC and bSQ structures are dynamically stable, while those in the HX structure are unstable. In this respect, most transition metals are classified into group A in Fig.~\ref{fig2}. Interestingly, the overall shape of the dispersion curves in the (12,6)-LJ crystals in the HX, bHC, and bSQ structures is quite similar to that in 2D Ni, Pd, and Pt (see Supplemental Material (SM) \cite{SM} including Fig.~20 in Ref.~\cite{ono2020_2}). The HCP Co may be classified into group B because the bHC structure is dynamically stable only (see SM \cite{SM} including Figs.~8 and 19 in Ref.~\cite{ono2020_2}). Recent DFT and empirical potential calculations have shown that alkali metals (Li, Na, and K) in the FCC, BCC, and HCP structures are dynamically stable \cite{takahashi,nichol}, while BCC Li is stabilized by anharmonic effects \cite{hellman}. For 2D Li, Na, and K, the bHC structure is dynamically stable only \cite{ono2020_2}. This indicates that there are no counterparts for the LJ crystals in group C, showing the limitation of the central potential for describing lattice dynamics of simple metals correctly. Although the overall shape of the phonon dispersions in the other alkali metals (Rb and Cs), alkali earth metals, and noble metals in the bHC and bSQ structures (see SM \cite{SM} including Figs.~11, 12, and 21 in Ref.~\cite{ono2020_2}) are similar to that in group B, the stability properties for either the HX, BCC, or HCP structures are different between the realistic metals and the LJ crystals: For Rb and Cs, BCC (HCP) structure is stable (unstable) \cite{takahashi}, while for the alkali earth and noble metals, the HX structure is dynamically stable. The boundary between the groups B, C, and D in Fig.~\ref{fig2} would be corrected by considering the many-body interaction beyond the central potential approximation \cite{finnis,foiles} or by considering the Friedel-like oscillation effect simply \cite{nichol}.  

\section{Conclusion}
We have studied the lattice dynamics of 2D and 3D LJ crystals by diagonalizing the dynamical matrix numerically and analytically. For all $(m,n)$, the HX structure is unstable due to the zero thickness. The buckling is important for obtaining dynamically stable 2D crystals: The bHC and/or bSQ structures are dynamically stable if the interaction between the LJ particles is described by the short-range potential. We have categorized the stability property into four groups shown in Fig.~\ref{fig2} and shown that the stability property of the LJ crystals with short-range potential ($m+n\ge 17$ and $n\ge 6$) is similar to that of transition metals in the periodic table. More parameters for the potential function are required to model the dynamical stability of metals in a unified manner. 


\appendix
\section{Derivation of Eq.~(\ref{eq:XbSQ})}
\label{app}
A straightforward calculation of the dynamical matrix at the point X yields 6$\times$6 matrix of $\tilde{D}_{ss'}^{\alpha\alpha'}(\bm{q}=(\pi/a_{\rm bSQ},0,0))$. By considering up to the second NN sites, the matrix elements are expressed as follows: For $s=s'$,
\begin{eqnarray}
\tilde{D}_{ss}^{xx} &=& 4\left[\tilde{A}+B(a_{\rm bSQ})+\alpha B(d_{12})\right],
\\ 
\tilde{D}_{ss}^{yy} &=& 4\left[\tilde{A} +\alpha B(d_{12})\right],
\\
\tilde{D}_{ss}^{zz} &=& 4\left[\tilde{A} +\beta B(d_{12})\right],
\end{eqnarray}
and for $s\ne s'$,
\begin{eqnarray}
\tilde{D}_{12}^{xz} &=& \tilde{D}_{12}^{zx} = -4iB(d_{12})\sqrt{\alpha\beta}, \\
\tilde{D}_{21}^{xz} &=& \tilde{D}_{21}^{zx} = 4iB(d_{12})\sqrt{\alpha\beta},
\end{eqnarray}
and the other elements are zero. The three eigenvalues (doubly degenerate) are obtained as
\begin{eqnarray}
 \omega_{1}^{2} &=& \frac{1}{2M}\left[ - \sqrt{(x - y)^2 + 4 z^2} + x + y\right],
 \\
 \omega_{2}^{2} &=& \frac{1}{2M}\left[ \sqrt{(x - y)^2 + 4 z^2} + x + y\right],
  \\
 \omega_{3}^{2} &=& \frac{y}{M}
\end{eqnarray}
with $x= \tilde{D}_{ss}^{xx}$, $y=\tilde{D}_{ss}^{yy}$, and $z= 4B(d_{12}) \sqrt{\alpha \beta}$. The condition $\omega_{1}^{2}>0$ is equivalent to Eq.~(\ref{eq:XbSQ}).



\begin{thebibliography}{99}

\bibitem{grimvall} G. Grimvall, B. Magyari-K\"ope, V. Ozoli\ifmmode \mbox{\c{n}}\else \c{n}\fi{}\ifmmode \check{s}\else \v{s}\fi{}, and K. A. Persson, Lattice instabilities in metallic elements, Rev. Mod. Phys. {\bf 84}, 945 (2012).

\bibitem{KS} W. Kohn and L. J. Sham, Self-Consistent Equations Including Exchange and Correlation Effects, Phys. Rev. {\bf 140}, A1133 (1965).

\bibitem{kawazoe} K. Parlinski, Z. Q. Li, and Y. Kawazoe, First-Principles Determination of the Soft Mode in Cubic ZrO$_2$, Phys. Rev. Lett. {\bf 78}, 4063 (1997).

\bibitem{dfpt} S. Baroni, S. Gironcoli, A. D. Corso, and P. Giannozzi, Phonons and related crystal properties from density-functional perturbation theory, Rev. Mod. Phys. {\bf 73}, 515 (2001).

\bibitem{togo} A. Togo and I. Tanaka, First principles phonon calculations in materials science, Scr. Mater. {\bf 108}, 1 (2015).

\bibitem{huang} X. Huang, S. Li, Y. Huang, S. Wu, X. Zhou, S. Li, C. L. Gan, F. Boey, C. A. Mirkin, and H. Zhang, Synthesis of hexagonal close-packed gold nanostructures, Nat. Commun. {\bf 2}, 292 (2011).

\bibitem{schonecker} S. Sch\"{o}necker, X. Li, M. Richter, and L. Vitos, Lattice dynamics and metastability of FCC metals in the HCP structure and the crucial role of spin-orbit coupling in platinum, Phys. Rev. B {\bf 97}, 224305 (2018).

\bibitem{ono2020_1} S. Ono, Two-dimensional square lattice polonium stabilized by the spin-orbit coupling, Sci. Rep. {\bf 10}, 11810 (2020).  

\bibitem{ono2020_2} S. Ono, Dynamical stability of two-dimensional metals in the periodic table, Phys. Rev. B {\bf 102}, 165424 (2020).  

\bibitem{mermin} N. W. Ashcroft, N. D. Mermin, and D. Wei, {\it Solid State Physics}, revised edition, (Cengage, Boston, 2016).

\bibitem{born} M. Born and K. Huang, {\it Dynamical Theory of Crystal Lattices}, (Oxford University Press, 1954).

\bibitem{wallace} D. C. Wallace and J. L. Patrick, Stability of Crystal Lattices, Phys. Rev. {\bf 137}, A152 (1965).  

\bibitem{mihalkovic} M. Mihalkovi\ifmmode \check{c}\else \v{c}\fi{} and C. L. Henley, Empirical oscillating potentials for alloys from ab initio fits and the prediction of quasicrystal-related structures in the Al-Cu-Sc system, Phys. Rev. B {\bf 85}, 092102 (2012).

\bibitem{alsalmi} O. Alsalmi, M. Sanati, R. C. Albers, T. Lookman, and A. Saxena, First-principles study of phase stability of bcc $X\mathrm{Zn}$ ($X=\mathrm{Cu}$, Ag, and Au) alloys, Phys. Rev. Materials {\bf 2}, 113601 (2018).

\bibitem{finnis} M. W. Finnis and J. E. Sinclair, A simple empirical $N$-body potential for transition
metals, Philos. Mag. A {\bf 50}, 45 (1984). 

\bibitem{foiles} S. M. Foiles, M. I. Baskes, and M. S. Daw, Embedded-atom-method functions for the fcc metals Cu, Ag, Au, Ni, Pd, Pt, and their alloys, Phys. Rev. B {\bf 33}, 7983 (1986).

\bibitem{ono2019_jap} S. Ono, Lattice dynamics for isochorically heated metals: A model study, J. Appl. Phys. {\bf 126}, 075113 (2019).

\bibitem{ono2020_jap} S. Ono and D. Kobayashi, Phonon softening in sodium with a stepwise electron distribution, J. Appl. Phys. {\bf 127}, 165105 (2020).

\bibitem{SM} See Supplemental Material at XXX for the phonon dispersion curves of elemental metals, extracted from Ref.~\cite{ono2020_2}.

\bibitem{takahashi} A. Takahashi, A. Seko, and I. Tanaka, Linearized machine-learning interatomic potentials for non-magnetic elemental metals: Limitation of pairwise descriptors and trend of predictive power, J. Chem. Phys. {\bf 148}, 234106 (2018).

\bibitem{nichol} A. Nichol and G. J. Ackland, Property trends in simple metals: An empirical potential approach, Phys. Rev. B {\bf 93}, 184101 (2016).

\bibitem{hellman} O. Hellman, I. A. Abrikosov, and S. I. Simak, Lattice dynamics of anharmonic solids from first principles, Phys. Rev. B {\bf 84}, 180301(R) (2011).

\end{thebibliography}
\end{document}